\documentclass[preprint,12pt,3p]{elsarticle}

\usepackage{array}
\newcolumntype{L}[1]{>{\raggedright\let\newline\\\arraybackslash\hspace{0pt}}p{#1}}
\newcolumntype{C}[1]{>{\centering\let\newline\\\arraybackslash\hspace{0pt}}p{#1}}
\newcolumntype{R}[1]{>{\raggedleft\let\newline\\\arraybackslash\hspace{0pt}}p{#1}}

\usepackage[bottom]{footmisc}

\usepackage{svg}

\usepackage{amssymb}

\usepackage{multirow}
\usepackage{multicol}
\usepackage{float}
\usepackage{amsthm}
\usepackage{amsmath}
\usepackage{rotating}

\usepackage{graphicx}
\usepackage{makecell}
\usepackage{tabularx}
\usepackage{xcolor,graphicx,float}  
\usepackage{multirow}
\usepackage{todonotes}
\usepackage{makecell}
\usepackage[hidelinks]{hyperref}
\usepackage{colortbl}
\usepackage{threeparttable}
\usepackage{booktabs} 

\usepackage{graphicx}
\usepackage{xcolor}
\usepackage{longtable}

\journal{Transportation Research Part A}

\begin{document}

\begin{frontmatter}

%\title{Modeling E-Bike Route Choice with Street-Level Imagery and Infrastructure Attributes: A Path Size Logit Analysis in Washington, DC}

\title{Modeling E-Bike Route Choice in Washington, DC: A Path Size Logit Approach}

\author[label1]{Yiheng Qian}
\ead{yihengqian@ufl.edu}

\author[label2]{Zehui Yin}
\ead{yinz39@mcmaster.ca}

\author[label2]{Darren M. Scott}
\ead{scottdm@mcmaster.ca}

\author[label1]{Xiang Yan\corref{cor1}}
\ead{xjacobyan@gmail.com}

\cortext[cor1]{Corresponding author}

\address[label1]{Department of Civil and Coastal Engineering, University of Florida, 1949 Stadium Rd, Gainesville, FL 32611, USA}

\address[label2]{TransLAB (Transportation Research Lab), School of Earth, Environment \& Society, McMaster University, 1280 Main Street West, Hamilton, ON L8S 4K1, Canada}

% \author{%
%   \textbf{Yiheng Qian}\\
%   Department of Civil and Coastal Engineering, University of Florida, 1949 Stadium Rd, Gainesville, FL 32611\\
%   yihengqian@ufl.edu\\
%    \hfill\break%
%   \textbf{Zehui Yin}\\
%   TransLAB (Transportation Research Lab), School of Earth, Environment \& Society, McMaster University, 1280 Main Street West, Hamilton, ON L8S 4K1\\
%   yinz39@mcmaster.ca\\
%   \hfill\break% this is a way to add line numbering on empty line
%   \textbf{Darren M. Scott}\\
%   TransLAB (Transportation Research Lab), School of Earth, Environment \& Society, McMaster University, 1280 Main Street West, Hamilton, ON L8S 4K1\\
%   scottdm@mcmaster.ca\\  
%   \hfill\break% this is a way to add line numbering on empty line
%   \textbf{Xiang Yan}\\
%   Department of Civil and Coastal Engineering, University of Florida, 1949 Stadium Rd, Gainesville, FL 32611\\
%   xiangyan@ufl.edu\\
%   \hfill\break
% }

\begin{abstract}
\textcolor{black}{
Understanding e-bike route choice is essential for developing effective cycling infrastructure, yet empirical evidence remains limited. This study investigates shared e-bike route choice in Washington, DC, using Global Positioning System (GPS) trajectory data from the Capital Bikeshare system. A Path Size Logit model is estimated using a hybrid choice set consisting of observed routes and corresponding shortest paths, integrating Geographic Information System (GIS)-based infrastructure variables with computer vision-derived street-level visual features extracted from Street View images (SVI). The results indicate that e-bike riders tend to choose routes that minimize conflicts with both motor vehicles and pedestrians while maintaining travel continuity. Roadway hierarchy substantially moderates the influence of bicycle facilities, with the presence of bicycle facilities having a much greater impact on route choice along major roads than along minor roads. Longer trips also exhibit stronger preferences for cycling infrastructure. Incorporating street-level visual features improves model performance, although their effects are generally smaller than those of road infrastructure, with trees being the only greenery component showing a consistently positive effect. Standardized effect sizes further identify the most behaviorally important route attributes. These findings provide practical evidence for cycling infrastructure planning in the e-bike era.
}
%This study examines electric bike (e-bike) route choice behavior in Washington, DC, using a Path Size Logit model estimated from GPS-based trip trajectory data. The analysis evaluates how route selection is associated with trip characteristics, cycling infrastructure, elevation, and street-level visual  features derived from Street View imagery. Results show that protected bike lanes substantially increase route attractiveness on both major and minor roads. Painted bike lanes display clear context dependence, performing poorly on major roads but showing a positive association on minor roads. Shared lanes strongly deter route choice on major roads, while showing no statistically significant effect on minor roads. Beyond on-road facilities, bikeable sidewalks and residential streets are positively associated with route choice. Street-level street-level visual  features also show systematic relationships with route choice, with sky visibility, walls, and the presence of cars associated with higher route attractiveness. By integrating micro-built environment information into a Path Size Logit framework, this study provides new empirical evidence on e-bike route choice in an American city context and demonstrates the value of combining behavioral modeling with computer-vision-derived environmental measures to support micromobility planning. Comparative evidence further indicates that e-bike route choice shares core sensitivities with bicycles and e-scooters, while also reflecting mode-specific responses related to travel efficiency and slope tolerance.

\end{abstract}

\begin{keyword}
%mobility-on-demand; public transit; traveler preferences; disadvantaged travelers; low-income community; ordered logit model
E-Bike \sep Route Choice \sep GPS \sep Street View Imagery \sep Path Size Logit
\end{keyword}

\end{frontmatter}

\section{Introduction}

%\citep{lilasathapornkit2025cycling}
%\citep{arning2023review}
%\citep{fishman2016bikes}

Electric bikes (e-bikes) have emerged as an increasingly important mode of urban transportation, offering a flexible and energy-efficient alternative for short- and medium-distance travel. As cities seek to promote low carbon mobility and reduce reliance on private automobiles, e-bikes are becoming an integral component of urban transport systems \citep{mcqueen2020bike}. Understanding how e-bike users select cycling routes is essential for designing effective cycling infrastructure, improving safety, and enhancing the overall riding experience \citep{dane2019route}. %Such knowledge also supports broader planning decisions, including the placement of supporting facilities such as bicycle parking and charging infrastructure.

Although a growing body of research has examined bicycle route choice behavior, empirical evidence focusing specifically on e-bike riding remains limited. Compared with conventional cyclists, e-bike users typically travel at higher speeds and are more capable of sharing the roadway with motor vehicles, potentially reducing their reliance on dedicated cycling infrastructure \citep{lopez2017unveiling}. Consistent with this expectation, several studies have found that e-bike users are less sensitive to differences among bicycle facility types than conventional cyclists \citep{lopez2017unveiling, khavarian2024bike, lilasathapornkit2025cycling}. However, other studies continue to report clear benefits of dedicated cycling infrastructure for e-bike route choice \citep{singh2025analysing, khavarian2024bike}. One possible explanation is that the effectiveness of bicycle facilities depends not only on the facility design itself but also on the roadway environment in which it is implemented. Evidence from bicycle research suggests that cyclists' perceptions of safety and comfort vary substantially across roadway contexts for the same facility type \citep{mcneil2015influence, monsere2014lessons, fosgerau2023bikeability}. Such context dependence may be even more pronounced for e-bike users because their higher operating speeds and different interactions with surrounding traffic can alter how infrastructure is perceived and utilized. Consequently, existing evidence on e-bike preferences for bicycle facilities remains inconclusive, highlighting the need for further quantitative evidence that explicitly distinguishes facility types across different roadway contexts.

Beyond infrastructure characteristics, the role of the micro-built environment in shaping e-bike route choice remains poorly understood. Existing route choice models primarily rely on Geographic Information System (GIS)-based infrastructure variables derived from OpenStreetMap (OSM) or municipal datasets \citep{cubells2023scooter,chung2024understanding,lilasathapornkit2025cycling,lukawska2023joint,meister2023route}. While these data effectively characterize the physical transportation network, they cannot directly capture the visual and perceptual qualities of the street environment. Features such as greenery, sky visibility, buildings, and other streetscape elements may influence perceived comfort, safety, and riding experience, yet have rarely been incorporated into route choice models because of the lack of scalable and systematic measurement approaches \citep{meng2023personalized}. As a result, it remains unclear whether these perceptual streetscape characteristics provide additional explanatory power beyond conventional GIS-based infrastructure measures in explaining e-bike route choice.

\begin{figure}[H]
  \centering
  \includegraphics[width=0.7\textwidth]{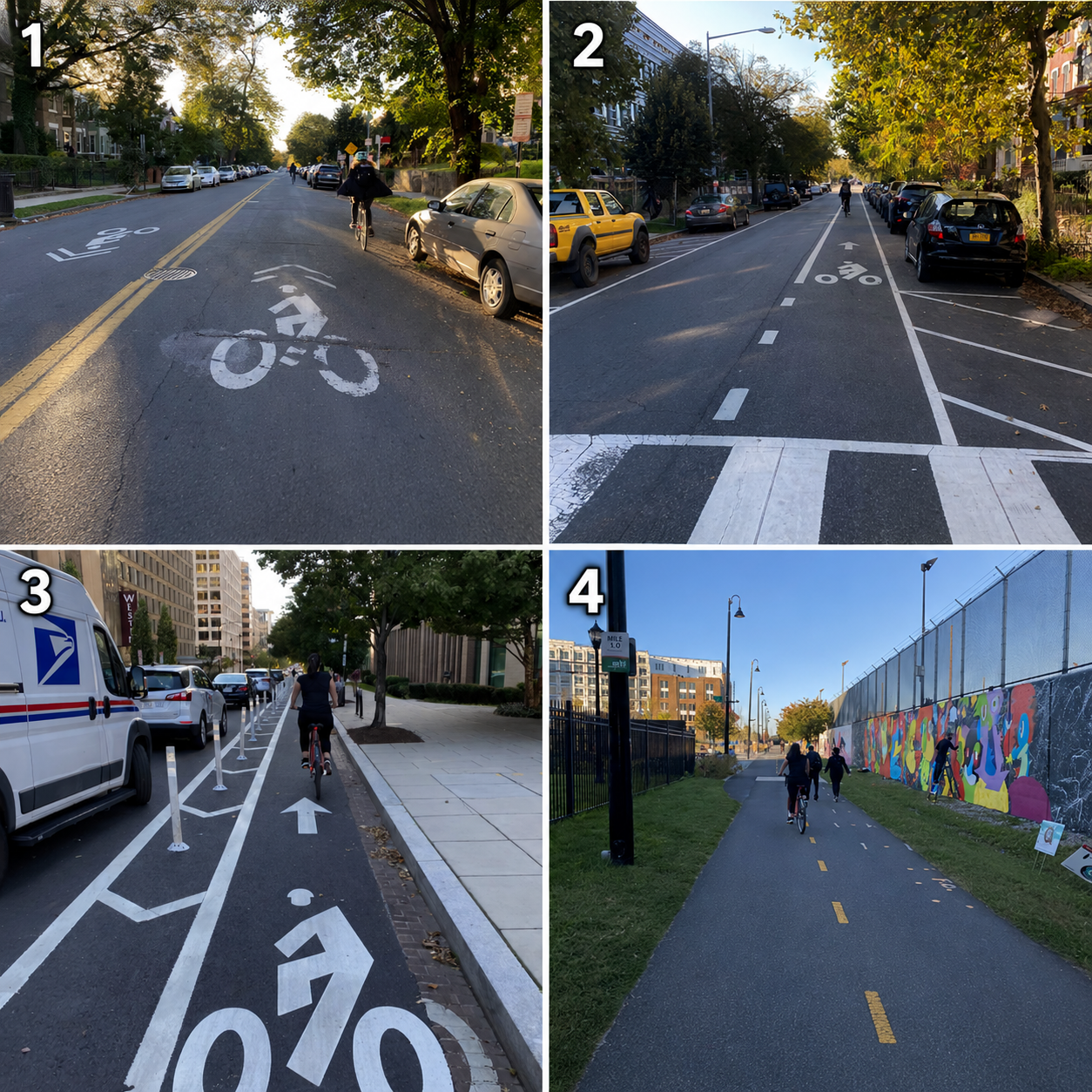}
  \caption{Illustration of bicycle facility types: 
  (1) Shared lane (sharrows) on R Street NE,
  (2) Painted lane on R Street NW, 
  (3) Protected lane on M Street NW, 
  (4) Shared-use path on MET Branch Trail.
  Source: District Department of Transportation. \citep{ddot2020bicycle}}
  \label{var1}
\end{figure}

Motivated by these research gaps, this study investigates shared e-bike route choice behavior in Washington, DC, which offers a diverse range of cycling environments (Figure \ref{var1}). The analysis is based on Global Positioning System (GPS) trajectory data from Capital Bikeshare, one of the largest docked bikeshare systems in the United States (U.S.). This study has two primary objectives. \textcolor{black}{First, it examines how different bicycle facility types influence e-bike route choice and whether these effects vary across roadway hierarchies.} Second, it quantifies the influence of street-level visual features, including trees, grass, plants, sky, buildings, and other streetscape elements. To address these objectives, we develop a Path Size Logit (PSL) model that integrates GIS-based infrastructure measures with computer vision-derived street-level environmental features extracted from Street View images (SVI). \textcolor{black}{This integrated framework enables a direct comparison of the relative influence of physical infrastructure and street-level visual environments on e-bike route choice, while quantifying the practical importance of actionable route attributes through standardized effect sizes.} To our knowledge, this study is the first revealed preference route choice analysis of shared e-bikes in a U.S. urban context. The findings provide large-scale empirical evidence on the relative influence of bicycle facilities, roadway hierarchy, and street-level visual environments on e-bike route choice, while demonstrating the value of integrating computer vision–derived streetscape measures into behavioral route choice models.

\section{Literature review}

\textcolor{black}{The following subsections review existing route choice modeling approaches and the key determinants of micromobility route choices, before identifying the remaining knowledge gaps addressed by this study.}

\subsection{Route choice modeling approaches}

\textcolor{black}{Route choice studies commonly rely on either stated preference (SP) or revealed preference (RP) data. SP methods, commonly implemented through surveys or hypothetical scenarios, are useful for capturing attitudes but may not fully capture how people actually travel. GPS trajectory data directly record realized travel behavior and are therefore well suited to route choice analysis \citep{fitch2020road,hsueh2023influential}. A range of modeling frameworks have been applied to GPS-based route choice data. Detour-based approaches quantify deviations from the shortest path as a measure of route preference \citep{park2019bicyclists,cubells2023scooter}. Link-based models, such as the Recursive Logit \citep{fosgerau2013link} and Perturbed Utility Route Choice \citep{fosgerau2022perturbed} models, assume that travelers make route choices through a sequence of link-level decisions, thereby avoiding explicit route choice set generation. In contrast, route-based discrete choice models treat entire routes as discrete alternatives and estimate attribute preferences within a random utility maximization framework. Among route-based approaches, the PSL model is widely used for bicycle route choice analysis \citep{ton2018evaluating,marra2020determining,nielsen2021relevance,yap2021taking,sevtsuk2021big,scott2021route,meister2023route}. It extends the Multinomial Logit framework by incorporating a path-size term to account for route overlap, thereby relaxing the independence assumption where competing routes share common segments \citep{bekhor2006evaluation,prato2009route}. Other extensions of the Logit framework have also been applied to route choice problems, including Mixed Logit models, which accommodate random taste heterogeneity and more flexible substitution patterns \citep{bernardi2018modelling}, and the recently proposed Smooth Bounded Choice Model, which integrates implicit choice set formation while retaining closed-form choice probabilities \citep{cazor2025smooth}.}

\textcolor{black}{The performance of route-based discrete choice models depends critically on the quality of the generated choice sets, making choice set construction a key methodological challenge. Existing micromobility studies have adopted a wide range of approaches, including using only the shortest path as the non-chosen alternative \citep{hsueh2023influential}, navigation-based route generation \citep{xiao2025effects}, shortest path–based algorithms such as the Metropolis--Hastings algorithm \citep{schumann2025city} and the Breadth First Search Link Elimination (BFS-LE) algorithm \citep{rieser2013route,ton2018evaluating}, as well as clustering observed routes when sufficient trajectory data are available for the same origin and destination (OD) pair \citep{cho2022estimation}. Choice set quality is commonly evaluated by its ability to reproduce observed routes while maintaining sufficient diversity among alternatives. Despite substantial methodological developments, no consensus has emerged regarding the optimal choice set generation strategy. Existing methods inevitably involve trade-offs among computational efficiency, behavioral realism, observed-route coverage, and route diversity \citep{felder2022choice,fitch2020road,halldorsdottir2014efficiency}.}

\subsection{Factors influencing route choice of micromobility}

Compared with e-bike route choice research, bicycle route choice has been studied much more extensively and provides a useful foundation for understanding e-bike behavior. A well-established body of literature has consistently shown that cyclists prefer shorter routes with dedicated cycling infrastructure \citep{lukawska2023joint}, while avoiding mixed traffic, intersections, and steep slopes \citep{meister2024comparative}. Beyond the presence of bicycle facilities, empirical studies have demonstrated substantial differences in the level of protection provided by different facility types \citep{nolan2021bicycle,mcneil2015influence,monsere2014lessons}. Route choice studies further indicate that protected facilities provide the greatest utility, particularly on major roads where separation from motor traffic is most beneficial \citep{fosgerau2023bikeability}. \textcolor{black}{In addition to physical infrastructure, route choice has also been linked to streetscape characteristics. Scenic environments, particularly those with abundant greenery, have consistently been associated with more attractive cycling routes \citep{lukawska2023joint,fosgerau2023bikeability}. At the micro-built environment level, greenery has received the greatest attention and has been shown to positively influence route choice \citep{park2019bicyclists,juarez2023cyclists,xiao2025effects,lieu2026comparing,lukawska2023joint}, cycling activity \citep{gao2021urban,wang2020relationship}, and bicycle mode choice \citep{lu2019associations,he2024choose}. Overall, existing evidence suggests that the contribution of street-level visual features is generally smaller than that of cycling infrastructure \citep{juarez2023cyclists,bialkova2022design,xiao2025effects}.}

These findings may not fully translate to e-bike users, as e-bikes differ from conventional bicycles in several important aspects. \textcolor{black}{Existing studies have shown that e-bike users typically travel longer distances and can access destinations that may be impractical using conventional bicycles \citep{hallberg2021modelling,bourne2020impact,lopez2017unveiling,plazier2017cycling,fyhri2015effects}. These changes may also alter trip purposes and the contexts in which routes are selected.} Moreover, although e-bikes offer higher travel speeds, these advantages are not uniform across the road network and may depend on roadway type and cycling infrastructure \citep{langford2015risky}. Together, these differences suggest that findings from conventional bicycle studies cannot be directly generalized to e-bike route choice, highlighting the need for dedicated empirical evidence.

\subsection{Summary of research gaps}

Empirical evidence on e-bike route choice remains limited, and existing studies report context-dependent and sometimes heterogeneous findings. A study in Dublin, Ireland, emphasized the importance of safe and uninterrupted cycling space, highlighting the role of protected infrastructure in encouraging e-bike use \citep{singh2025analysing}. In contrast, a study conducted in Helsinki, Finland, suggested that e-bike users were relatively less sensitive to facility type and road classification, although dedicated cycling routes remained beneficial \citep{khavarian2024bike}, while no statistically significant preference for protected infrastructure was observed among e-bike riders in Melbourne, Australia \citep{lilasathapornkit2025cycling}. \textcolor{black}{In Flanders, Belgium, e-bike users also appeared more willing to share the roadway with other traffic participants than conventional cyclists \citep{lopez2017unveiling}.} Beyond infrastructure, an interview study conducted in Groningen, the Netherlands, found that e-bike users generally preferred enjoyable and quiet routes except when constrained by time or adverse weather \citep{plazier2017cycling}. \textcolor{black}{At the streetscape level, a study in Jinan, China, reported that overhead greenery and open sky index were negatively associated with e-bike route choice \citep{yu2022exploring}.}

The discrepancies in the existing literature highlight several important knowledge gaps. First, despite the rapid growth of shared e-bike systems, revealed preference route choice studies remain limited, particularly in North American urban environments. Second, compared with the broader bicycle route choice literature, existing e-bike studies have primarily focused on broad infrastructure categories, while the effects of different cycling facility types across diverse roadway hierarchies remain insufficiently understood. This question is especially relevant for e-bike users, who are more likely to travel longer distances and use a wider range of road types than conventional cyclists. Third, although streetscape characteristics have received increasing attention, most studies have focused on greenery, with limited evidence regarding the influence and relative contribution of a broader set of street-level visual features compared with conventional infrastructure variables.

This study addresses these gaps by examining shared e-bike route choice in Washington, DC using GPS-based revealed preference data from the Capital Bikeshare system. We apply a PSL model to investigate how cycling infrastructure, roadway hierarchy, and Street View-derived street-level visual features influence route choice. By integrating GIS-based infrastructure variables with computer vision–derived streetscape measures, this study provides new evidence on the relative importance of physical and visual route attributes for e-bike route choice in a North American urban context, offering implications for cycling infrastructure planning and network design.

\section{Data}
\subsection{E-bike trip trajectories}

\begin{figure}[!t]
  \centering
  \includegraphics[width=1\textwidth]{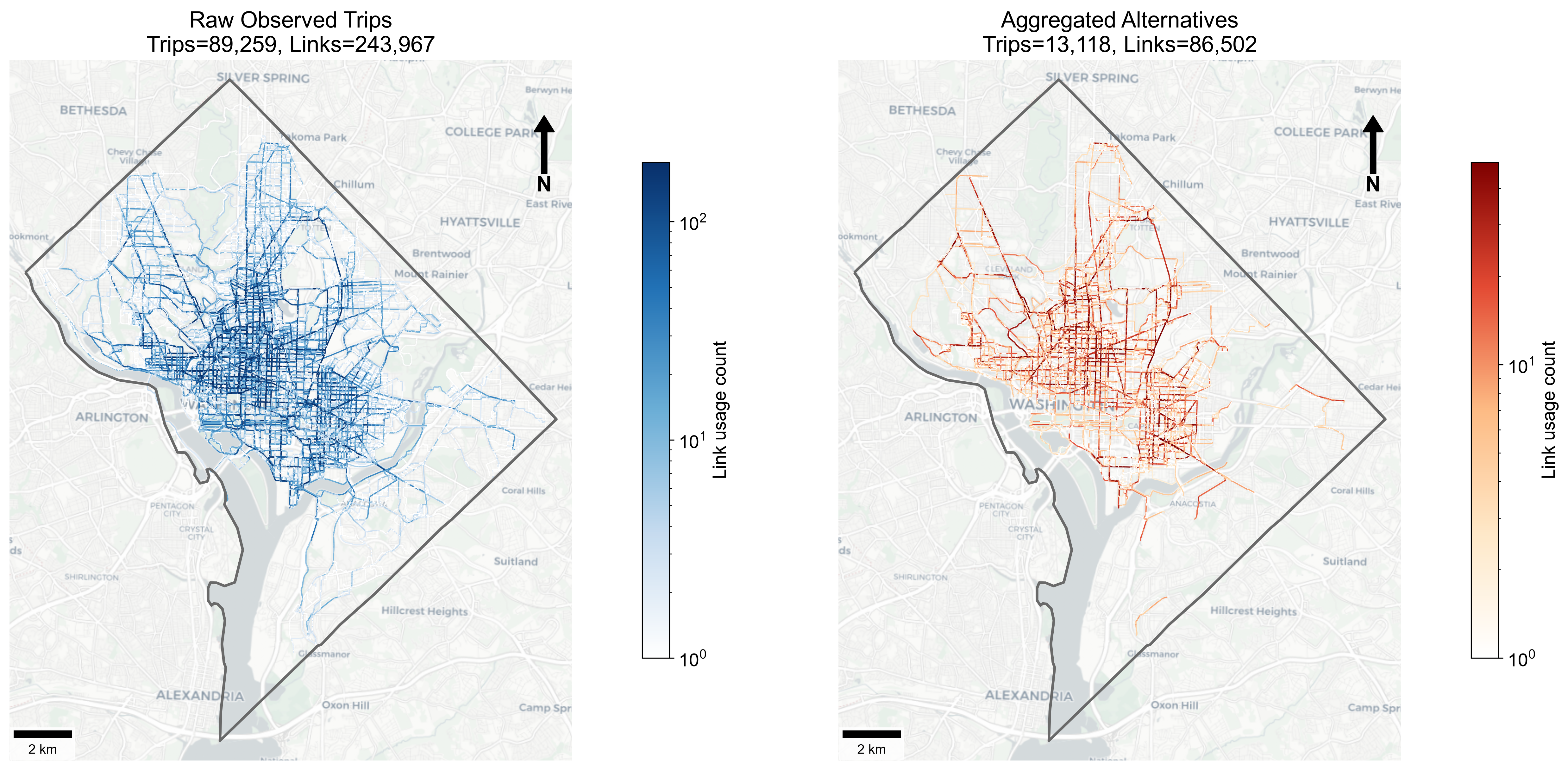}
  \caption{Study area and e-bike route data.}
  \label{area}
\end{figure}

This study focuses on Washington, DC, where bikesharing was first introduced through SmartBike DC in 2008, followed by the launch of Capital Bikeshare in 2010. Electric-assist bicycles were subsequently introduced through a pilot deployment within the Capital Bikeshare system in 2018 \citep{DC_CaBi_Ebike_2018}. Capital Bikeshare is now operated by Lyft. We use GPS trajectory data from shared e-bikes provided by Lyft, covering the entire calendar year of 2023. Each entry in the dataset represents a GPS point, including latitude, longitude, a timestamp, and a unique e-bike ID that remains stable over time. This consistent ID allows us to accurately reconstruct individual e-bike trajectories. To ensure data quality, we refined the trip data by excluding trips shorter than 2 minutes, longer than 90 minutes, \textcolor{black}{or distances outside 600 to 12000 meters. After processing, 89,259 valid trips were identified for this study, covering all trip trajectories in DC. Figure~\ref{area} shows the study area along with the e-bike data.}

%The cleaned and structured trip dataset forms the basis for our subsequent modeling of e-bike route choice behavior. For this, we implemented a three-step methodological framework: 1) preparing the road network and applying a map matching algorithm to align raw GPS trajectories with the transport infrastructure; 2) cleaning and refining matched trajectories to ensure valid origin-destination (OD) pairs; and 3) constructing behaviorally plausible choice sets for route choice modeling. These steps ensure that the input data are spatially accurate and behaviorally meaningful for subsequent model estimation.

\subsection{Network and map matching}

Map matching is a crucial step that assigns GPS-based trajectories to a series of links or nodes (and related network attributes) to precisely reconstruct users' actual paths. As reviewed in a recent study by \citet{berjisian2023evaluation}, map matching for active travel modes is more complex than for car travel, highlighting two main challenges:
1) Cyclists and pedestrians do not always stick to official roads and paths, sometimes opting for shortcuts that may be missing from the network.
2) The types of infrastructure these users are allowed to use, and actually do use, are not always clearly defined. Individuals often switch between close parallel facilities that traditional matching algorithms struggle to detect.

We used the Python package Leuven.MapMatching \citep{meert2018hmm} to map-match the trip GPS trajectories based on a Hidden Markov Model with non-emitting states \citep{newson2009hidden}. The map-matching was conducted on the OSM walking road network \citep{OpenStreetMap}, with one-way street restrictions removed to better reflect the fewer constraints typically faced by cyclists compared to drivers. \textcolor{black}{The resulting network comprised 652,280 edges and 291,612 nodes, with an average edge length of 18 m. After map matching, 243,967 edges (37.4\% of the network) were traversed by at least one observed trip, indicating broad spatial coverage across the study area.}

To enhance matching accuracy, we applied a series of custom preprocessing steps, including the removal of GPS outliers and dwelling points. Additionally, turn events were extracted during the map-matching process by analyzing changes in road segment bearings. Specifically, bearing changes between 75° and 170° were classified as right turns, and between 190° and 285° as left turns. This map-matching process was first introduced and used in the study by \citet{yin2026decoding}. Using the selected valid trip trajectories and the OSM walking road network \citep{OpenStreetMap}, we applied customized map-matching, assigning each trip a unique ID. The output included the trip ID, the sequence of traversed links and nodes, their corresponding geometries matched to the road network, and the counts of left and right turns. We manually verified a subset of randomly sampled routes by comparing the map-matched paths to the original GPS trajectories. Manual inspection of a randomly sampled subset showed that the map-matched routes closely mirrored the original GPS trajectories. To prepare inputs for the choice set generation, we conducted post-processing on the map-matching results. Trips containing duplicate nodes or links within a single trajectory, such as circular routes or back-and-forth movements, were excluded. These patterns often indicate non-purposeful or leisure travel rather than trips between distinct locations. In many such cases, the OD are very close, resulting in unrealistically short alternative routes that do not reflect the user's actual decision-making. Including these trips could compromise the validity of route choice modeling.

\subsection{Street View images}

\begin{figure}[H]
  \centering
  \includegraphics[width=1\textwidth]{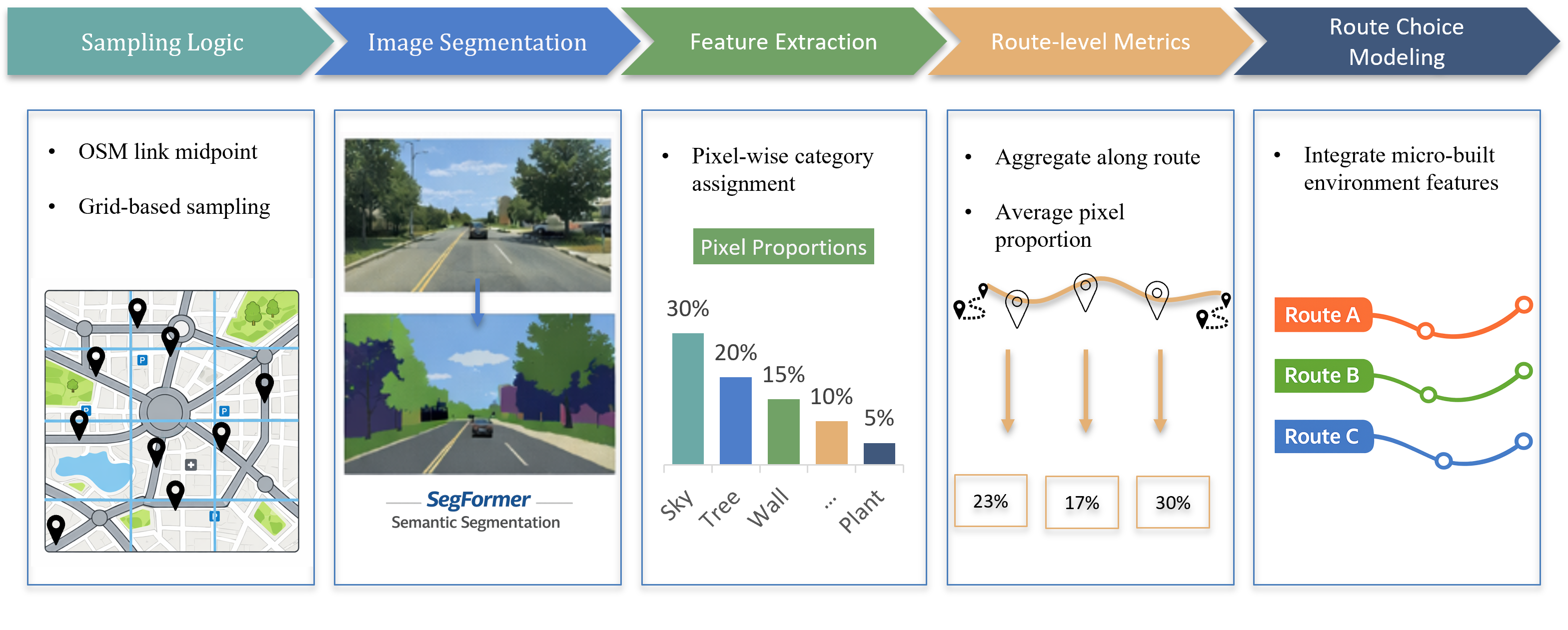}
  \caption{Street View image processing.}
  \label{gsv}
\end{figure}

To capture the visual and perceptual qualities of the riding environment, we extracted street-level visual features from SVI using computer vision. A total of 54,601 images were sampled and collected from an online third-party provider at unique geographic locations along the candidate network. Each image was processed using SegFormer \citep{DBLP:journals/corr/abs-2105-15203}, a transformer-based semantic segmentation model pre-trained on ImageNet and fine-tuned on ADE20K. The model assigns pixel-level semantic labels to each image, quantifying the proportions of natural elements (e.g., tree, sky, grass) and built environment features (e.g., building, wall). The overall processing workflow is illustrated in Figure~\ref{gsv}.

{\footnotesize
\renewcommand\cellalign{tl}
\renewcommand\arraystretch{1.1}

\begin{longtable}{
>{\raggedright\arraybackslash}p{0.21\textwidth}
>{\raggedright\arraybackslash}p{0.71\textwidth}
}
\caption{\textcolor{black}{Variables in the e-bike route choice model}}\label{var}\\
\toprule
\textbf{Variable} & \textbf{Description} \\
\midrule
\endfirsthead

\multicolumn{2}{c}{\tablename\ \thetable\ (continued)}\\
\toprule
\textbf{Variable} & \textbf{Description} \\
\midrule
\endhead

\midrule
\multicolumn{2}{r}{Continued on next page}\\
\endfoot

\bottomrule
\endlastfoot
\multicolumn{2}{l}{\textbf{Trip characteristics}}\\
Length & Total length of all links comprising the route. \\
Turns & Total number of left and right turns. \\
\midrule

\multicolumn{2}{l}{\textbf{Elevation}}\\
Slope 2\%--6\% & Proportion of route length with 2\%--6\% positive slope. \\
Slope 6\%--10\% & Proportion of route length with 6\%--10\% positive slope. \\
Slope $>$10\% & Proportion of route length with slope greater than 10\%. \\
\midrule

\multicolumn{2}{l}{\textbf{Road infrastructure}}\\

\textit{Major} &
\emph{Definition:} OSM tag: highway = \{``primary'', ``secondary'', ``tertiary'', ``unclassified''\}, with at least two motor vehicle lanes in at least one direction. \\

Major w/ protected &
Proportion of route length on major roads with protected bike lanes. OSM tag: cycleway=\{``track'', ``separate''\}. \\

Major w/ painted &
Proportion of route length on major roads with painted bike lanes. OSM tag: cycleway=\{``lane'', ``opposite\_lane''\}. \\

Major w/ sharrows &
Proportion of route length on major roads with shared bike markings. OSM tag: cycleway=\{``shared\_lane'', ``share\_busway''\}. \\

Major w/o bike lane &
Proportion of route length on major roads without bicycle facilities. \\

\addlinespace

\textit{Minor} &
\emph{Definition:} OSM tag: highway = \{``primary'', ``secondary'', ``tertiary'', ``unclassified''\}, with at most one motor vehicle lane per direction. \\

Minor w/ protected &
Proportion of route length on minor roads with protected bike lanes. OSM tag: cycleway=\{``track'', ``separate''\}. \\

Minor w/ painted &
Proportion of route length on minor roads with painted bike lanes. OSM tag: cycleway=\{``lane'', ``opposite\_lane''\}. \\

Minor w/ sharrows &
Proportion of route length on minor roads with shared bike markings. OSM tag: cycleway=\{``shared\_lane'', ``share\_busway''\}. \\

Minor w/o bike lane &
Proportion of route length on minor roads without bicycle facilities. \\

\addlinespace

Residential &
Proportion of route length on residential roads.
OSM tag: highway=``residential''. \\

Crossings &
Proportion of route length on crossings.
OSM tag: highway=``footway'', footway=``crossing''. \\

Footpath &
Proportion of route length on footpaths.
OSM tag: highway=``footway'', excluding footways adjacent to parallel roads. \\

Off-street bike paths &
Proportion of route length on off-street cycleways.
OSM tag: highway=``cycleway''. \\

Pedestrian zones &
Proportion of route length on roads used mainly by pedestrians.
OSM tag: highway=``pedestrian''. \\

Living streets &
Proportion of route length on living streets.
OSM tag: highway=``living\_street''. \\

Shared paths &
Proportion of route length on shared paths between cyclists and pedestrians.
OSM tag: highway=\{``path'', ``service''\}. \\

Stairs &
Proportion of route length on flights of steps.
OSM tag: highway=``steps''. \\

\multicolumn{2}{l}{\textbf{Street-level visual features}}\\

Tree & Average proportion of tree pixels from SVI points along the route   \\
Plant & Average proportion of plant pixels from SVI points along the route   \\
Grass & Average proportion of grass pixels from SVI points along the route   \\
Sky & Average proportion of sky pixels from SVI points along the route   \\
Wall & Average proportion of wall pixels from SVI points along the route   \\
Building & Average proportion of building pixels from SVI points along the route  \\
Sidewalk & Average proportion of sidewalk pixels from SVI points along the route   \\

\midrule

$\ln(\mathrm{PS})$ & Path size term to account for route overlap  \\

\end{longtable}
}

\section{Route choice model}

This section presents the methodology for modeling e-bike route choice behavior. It first describes the explanatory variables included in the analysis, then introduces the PSL model, a discrete choice framework that captures route choice behavior by relating route selection to route attributes while explicitly accounting for overlap among alternative routes. Finally, it describes the construction of the hybrid choice sets used for model estimation.

\subsection{Variables affecting e-bike route choice}

\begin{table}[!t]
\centering
\footnotesize
\renewcommand\arraystretch{1.2}
\caption{\textcolor{black}{Descriptive statistics of observed routes and shortest paths}}
\label{desc-observed-shortest}
\vspace{4pt}

\begin{tabular}{lrrrr}
\hline
& \multicolumn{2}{c}{Observed routes}
& \multicolumn{2}{c}{Shortest paths} \\
\cmidrule(lr){2-3}\cmidrule(lr){4-5}
Variable & Mean & SD & Mean & SD \\
\hline

\multicolumn{5}{l}{\textbf{Trip characteristics}} \\

Length & 1.46 & 0.68 & 1.39 & 0.64 \\
Turns & 9.76 & 5.46 & 14.84 & 7.04 \\

\hline
\multicolumn{5}{l}{\textbf{Elevation}} \\

Slope 2\%-6\% & 13.40 & 10.07 & 12.81 & 9.94 \\
Slope 6\%-10\% & 5.89 & 6.58 & 5.70 & 6.45 \\
Slope $>$10\% & 8.11 & 9.27 & 8.51 & 8.86 \\

\hline
\multicolumn{5}{l}{\textbf{Road infrastructure}} \\

Major w/ protected & 4.17 & 11.31 & 3.98 & 11.53 \\
Major w/ painted & 3.85 & 10.11 & 3.75 & 10.68 \\
Major w/ sharrows & 2.10 & 7.97 & 1.92 & 7.73 \\
Major w/o bike lane & 10.24 & 21.63 & 12.23 & 23.20 \\
Minor w/ protected & 4.09 & 11.25 & 3.28 & 9.67 \\
Minor w/ painted & 18.92 & 26.08 & 17.34 & 25.21 \\
Minor w/ sharrows & 6.74 & 13.58 & 4.64 & 11.40 \\
Minor w/o bike lane & 14.20 & 21.83 & 14.39 & 22.97 \\
Residential & 18.97 & 25.89 & 18.24 & 25.27 \\
Crossings & 4.76 & 3.01 & 6.16 & 4.24 \\
Footpath & 2.28 & 6.21 & 4.28 & 8.16 \\

Off-street bike paths & 6.37 & 15.03 & 5.83 & 16.25 \\
Pedestrian zones & 0.08 & 0.73 & 0.12 & 0.90 \\
Living streets & 0.06 & 1.12 & 0.01 & 0.31 \\
Shared paths & 2.33 & 5.01 & 2.54 & 6.79 \\
Stairs & 0.01 & 0.08 & 0.04 & 0.22 \\

\hline
\multicolumn{5}{l}{\textbf{Street-level visual features}} \\

Tree & 14.21 & 5.36 & 13.97 & 5.53 \\
Plant & 1.03 & 0.65 & 1.01 & 0.64 \\
Grass & 1.00 & 1.21 & 0.99 & 1.20 \\
Sky & 22.44 & 3.96 & 22.59 & 3.98 \\
Wall & 0.79 & 0.88 & 0.81 & 0.95 \\
Building & 14.02 & 6.56 & 14.47 & 6.69 \\
Sidewalk & 5.46 & 1.73 & 5.59 & 1.71 \\
\hline
$\ln(\mathrm{PS})$ & -1.08 & 0.53 & -1.49 & 0.63 \\

\hline
\end{tabular}

\par\vspace{2mm}
\footnotesize
\textbf{Note:} Percentage variables are reported in percentage points.
\end{table}

To identify the factors that may influence e-bike users' route choices, we generated attributes for all routes within the choice set. We have considered four categories of variables: trip characteristics, elevation, road infrastructure, and street-level visual features derived from SVI, Table \ref{var} provides detailed information about these variables. 

The variable ``Turns'' captures both left and right turns along the route, calculated during the map-matching procedure. While a potential extension could involve separating left and right turns, we observed a high correlation between the two, so we retained a combined variable to avoid multicollinearity.

All road infrastructure and elevation variables were measured in proportions. The classification of road type and cycling infrastructure follows previous studies \citep{fosgerau2023bikeability,lukawska2023joint}, and we further refined the infrastructure categories in the context of e-bike travel. Specifically, we relied on OSM data, using the highway and cycleway tags to define road types and infrastructure combinations, as detailed in Table \ref{var}. Elevation attributes were derived from 2 foot contour elevation data available via Open Data DC \citep{DCGIS_TopographySpotElevations}. For each segment, slope was estimated by taking the elevation difference between its OD points, divided by the segment length. Informed by prior literature \citep{meister2023route}, slopes were then categorized into three groups: 2\%--6\%, 6\%--10\%, and greater than 10\%. \textcolor{black}{For the micro-built environment variables, we first identified the nearest SVI point to the midpoint of each edge. The average matching distance was 19 m, and the 95th percentile was 114 m. The attributes of the nearest SVI point were then assigned to each edge, and route-level feature values were calculated as length-weighted averages across all edges along each route (Figure~\ref{gsv}). This approach incorporates street-level visual context into the route choice model, complementing conventional GIS-based infrastructure variables and providing a more comprehensive representation of the riding environment.}

In the OSM road network, ``footway'' segments typically exist as independent paths, with some running parallel to adjacent main roads. Due to GPS uncertainty and the close proximity of sidewalks and bicycle facilities to adjacent roadways, it is often difficult to determine whether a matched trajectory truly represents riding on a sidewalk. This may lead to misclassification during map matching and consequently bias the estimated effects of roadway characteristics. \textcolor{black}{To better represent the overall riding environment, footway segments identified as parallel sidewalks were assigned the roadway classification of their adjacent roads. The remaining footway segments, which primarily represent off-street pedestrian facilities independent of the roadway network, were retained as a separate category and labeled as ``footpath''.} Similarly, OSM ``cycleway'' segments were labeled as off-street cycleways to distinguish them from on-road bicycle facilities.

\subsection{Path size logit model}

\textcolor{black}{Because route alternatives within the constructed choice sets frequently share common links, the PSL model was adopted as it explicitly accounts for route overlap within a route-based discrete choice framework.}

To calculate the path size term (\(PS_i\)) for route \(i\), we use the following formula:
\begin{equation}
    PS_i = \sum_{a \in i} \frac{L_a}{L_i} \frac{1}{\sum_{j \in C} \delta_{aj}}
\end{equation}
where \(L_a\) is the length of road link \(a\), \(L_i\) is the length of route \(i\), \(\delta_{aj}\) is an indicator function that equals 1 if road link \(a\) is in route \(j\) and 0 otherwise, and \(C\) is the choice set of all routes.

% Path Size Logit Model Utility
The latent deterministic utility (\(V_i\)) for route \(i\) in the PSL model is given by:
\begin{equation}
    V_i = \beta X_i + \ln(PS_i)
\end{equation}
where \(\beta\) is a vector of generic coefficients, \(X_i\) is a vector of observed attributes for route \(i\), and \(PS_i\) is the path size term for route \(i\).

% Probability of Choosing Route i
The probability (\(P_i\)) of choosing route \(i\) is then given by the standard multinomial logit formulation:
\begin{equation}
    P_i = \frac{e^{V_i}}{\sum_{j \in C} e^{V_j}}
\end{equation}

These equations together form the basis of the PSL model, which accounts for route overlap by incorporating the path size term into the utility calculation, thereby providing a more robust representation of route choice behavior.

\subsection{Choice set generation}

Motivated by the considerations discussed in the literature, we constructed a hybrid choice set that primarily relies on observed routes while incorporating the shortest path as a benchmark alternative.

\textcolor{black}{The hybrid choice set was constructed in four steps.}

\textcolor{black}{\textbf{Step 1. Origin-destination aggregation.} Trips with identical OD pairs were grouped into a common choice set following previous route choice studies \citep{ton2018evaluating,scott2021route}. To account for GPS uncertainty, origins and destinations were spatially clustered using a 150 m grid, such that trips whose origins and destinations fell within the same grid cells were treated as belonging to the same OD pair. This spatial tolerance corresponds to an average endpoint displacement of approximately 57 m\footnote{For a square grid with side length \(L\), the expected Euclidean distance from a uniformly distributed point to the cell center is
\[
E[d]=\frac{4}{L^2}\int_{0}^{L/2}\int_{0}^{L/2}\sqrt{x^2+y^2}\,dy\,dx
=0.3826L.
\]
For \(L=150\) m, this corresponds to an average positional displacement of approximately \(57\) m.
}. To ensure sufficient behavioral observations, only OD groups containing at least three observed trips were retained. After this step, 42,255 trips remained, representing approximately half of the original dataset.}

\textcolor{black}{\textbf{Step 2. Aggregation of highly similar observed routes.} The remaining trips could not be regarded as unique route alternatives because many differed only by minor map-matching variations. In our road network, the average edge length is approximately 18 m, meaning that even small positional differences can produce different link sequences. Therefore, observed routes were aggregated whenever at least 80\% of the shorter route overlapped with another route. After aggregation, the average overlap among routes within each aggregated alternative was 89\%, indicating that the grouped trajectories were highly similar given the expected map-matching uncertainty and OD aggregation tolerance. The attributes of each aggregated alternative were calculated as the average of the constituent observed trips.}

\textcolor{black}{\textbf{Step 3. Choice set filtering and shortest-path augmentation.} To improve modeling reliability, only OD groups in which the most frequently chosen observed alternative was selected at least twice were retained, reducing the influence of infrequently observed routes \citep{lu2018understanding}. This filtering yielded 13,118 observed trips distributed across 1,672 unique choice sets. Figure~\ref{area} shows that the filtered trips retained a spatial distribution broadly similar to that of the original observed trajectories. Because the true consideration set is unobserved in revealed-preference route choice analysis, relying only on realized routes may produce an incomplete and selection-biased choice set. We therefore augment each OD choice set with the network shortest path as a non-chosen alternative. The shortest path is a behaviorally plausible benchmark representing the minimum-distance route that was feasible but not selected. Including this reference alternative helps capture deviations from pure travel minimization and allows the model to identify the extent to which infrastructure, topography, and streetscape attributes explain observed route choices. Approximately 2.8\% of the generated shortest paths were identical to an existing observed alternative.}

\textcolor{black}{\textbf{Step 4. Expansion into discrete choice observations.} Finally, each aggregated route alternative was expanded according to its observed frequency. Within each expanded choice occasion, one route was coded as the chosen alternative (\texttt{chosen = 1}), while all remaining alternatives, including the shortest path, were coded as non-chosen (\texttt{chosen = 0}). The resulting dataset contained 13,118 choice occasions, equal to the total number of retained observed trips, as each observed trip corresponds to one choice occasion in which its aggregated alternative was selected. Table \ref{desc-observed-shortest} provides descriptive statistics of observed routes and shortest paths}

\textcolor{black}{The resulting hybrid choice set combines behaviorally observed alternatives with a representative shortest path, balancing behavioral realism with choice set completeness. This behavior-based choice set construction approach offers several advantages. First, because the candidate routes are primarily derived from observed route choices rather than algorithmically generated paths, the choice sets better reflect travelers' actual preferences, constraints, and route awareness, while requiring the shortest paths as benchmark alternatives. Second, aggregating highly overlapping observed routes reduces redundancy within each choice set while preserving behaviorally equivalent routes. Third, by retaining the observed frequencies of aggregated routes, the framework preserves information on the relative popularity of different alternatives, allowing route choice behavior to be modeled under naturalistic conditions.}

\section{Results}
\textcolor{black}{
The following sections present the estimated route choice model, examine context dependence through interaction effects, and compare the practical importance of route attributes using standardized effect sizes.}

\subsection{Model outputs}

\begin{table}[!t]
\centering
\footnotesize
\renewcommand\arraystretch{1.2}
\caption{\textcolor{black}{Path size logit model results}}
\label{results-1234}
\vspace{4pt}

\resizebox{\textwidth}{!}{%
\begin{tabular}{lcccccccc}
\hline
& \multicolumn{2}{c}{Model 1 (Base)} 
& \multicolumn{2}{c}{Model 2 (SVI)} 
& \multicolumn{2}{c}{Model 3 (Infra)} 
& \multicolumn{2}{c}{Model 4 (Full)} \\
\cmidrule(lr){2-3}\cmidrule(lr){4-5}\cmidrule(lr){6-7}\cmidrule(lr){8-9}
Variable & Coef. (SE) & z & Coef. (SE) & z & Coef. (SE) & z & Coef. (SE) & z \\
\hline

\multicolumn{9}{l}{\textbf{Trip characteristics}} \\

Length
& $0.64^{***}$ $(0.13)$ & $4.78$
& $0.64^{***}$ $(0.14)$ & $4.72$
& $-0.46^{**}$ $(0.16)$ & $-2.91$
& $-0.48^{**}$ $(0.16)$ & $-3.00$ \\

Turns
& $-0.31^{***}$ $(0.00)$ & $-69.04$
& $-0.31^{***}$ $(0.00)$ & $-68.85$
& $-0.28^{***}$ $(0.00)$ & $-59.86$
& $-0.28^{***}$ $(0.00)$ & $-59.52$ \\

\hline
\multicolumn{9}{l}{\textbf{Elevation}} \\

Slope 2\%-6\%
& $0.36$ $(0.23)$ & $1.54$
& $0.47^{*}$ $(0.23)$ & $2.03$
& $0.22$ $(0.24)$ & $0.91$
& $0.26$ $(0.24)$ & $1.08$ \\

Slope 6\%-10\%
& $-0.99^{**}$ $(0.34)$ & $-2.93$
& $-0.98^{**}$ $(0.34)$ & $-2.89$
& $-0.36$ $(0.35)$ & $-1.05$
& $-0.32$ $(0.35)$ & $-0.92$ \\

Slope $>$10\%
& $-1.84^{***}$ $(0.40)$ & $-4.62$
& $-2.29^{***}$ $(0.39)$ & $-5.87$
& $-1.59^{***}$ $(0.43)$ & $-3.74$
& $-1.96^{***}$ $(0.42)$ & $-4.71$ \\

\hline
\multicolumn{9}{l}{\textbf{Road infrastructure}} \\

Residential (ref.)
&  & 
&  & 
& -- & --
& -- & -- \\

Major w/ protected
&  & 
&  & 
& $2.42^{***}$ $(0.29)$ & $8.41$
& $2.71^{***}$ $(0.30)$ & $9.13$ \\

Major w/ painted
&  & 
&  & 
& $1.12^{*}$ $(0.52)$ & $2.16$
& $1.89^{***}$ $(0.52)$ & $3.61$ \\

Major w/ sharrows
&  & 
&  & 
& $1.25^{**}$ $(0.42)$ & $3.01$
& $1.71^{***}$ $(0.42)$ & $4.05$ \\

Major w/o bike lane
&  & 
&  & 
& $-1.93^{***}$ $(0.18)$ & $-10.57$
& $-1.43^{***}$ $(0.20)$ & $-7.25$ \\

Minor w/ protected
&  & 
&  & 
& $2.02^{***}$ $(0.34)$ & $5.86$
& $2.01^{***}$ $(0.34)$ & $5.84$ \\

Minor w/ painted
&  & 
&  & 
& $1.31^{***}$ $(0.14)$ & $9.39$
& $1.41^{***}$ $(0.14)$ & $9.84$ \\

Minor w/ sharrows
&  & 
&  & 
& $0.18$ $(0.22)$ & $0.80$
& $0.80^{***}$ $(0.23)$ & $3.48$ \\

Minor w/o bike lane
&  & 
&  & 
& $-0.27^{*}$ $(0.14)$ & $-1.98$
& $-0.26^{.}$ $(0.14)$ & $-1.90$ \\

Crossings
&  & 
&  & 
& $-9.05^{***}$ $(0.58)$ & $-15.64$
& $-8.92^{***}$ $(0.59)$ & $-15.24$ \\

Footpath
&  & 
&  & 
& $-1.88^{***}$ $(0.33)$ & $-5.66$
& $-1.62^{***}$ $(0.34)$ & $-4.76$ \\

Off-street bike paths
&  & 
&  & 
& $1.81^{***}$ $(0.20)$ & $9.26$
& $2.05^{***}$ $(0.20)$ & $10.24$ \\

Pedestrian zones
&  & 
&  & 
& $-6.77^{*}$ $(2.69)$ & $-2.52$
& $-6.45^{*}$ $(2.71)$ & $-2.38$ \\

Living streets
&  & 
&  & 
& $0.66$ $(1.46)$ & $0.45$
& $0.49$ $(1.49)$ & $0.33$ \\

Shared paths
&  & 
&  & 
& $-2.39^{***}$ $(0.40)$ & $-6.02$
& $-2.26^{***}$ $(0.40)$ & $-5.59$ \\

Stairs
&  & 
&  & 
& $-130.20^{***}$ $(20.45)$ & $-6.37$
& $-139.72^{***}$ $(21.05)$ & $-6.64$ \\

\hline
\multicolumn{9}{l}{\textbf{Street-level visual features}} \\

Tree
&  & 
& $3.57^{***}$ $(0.78)$ & $4.58$
&  & 
& $3.27^{***}$ $(0.86)$ & $3.82$ \\

Plant
&  & 
& $6.17$ $(4.71)$ & $1.31$
&  & 
& $5.76$ $(5.11)$ & $1.13$ \\

Grass
&  & 
& $-4.83$ $(3.38)$ & $-1.43$
&  & 
& $5.82$ $(3.79)$ & $1.54$ \\

Sky
&  & 
& $-1.20$ $(0.80)$ & $-1.50$
&  & 
& $-3.32^{***}$ $(0.88)$ & $-3.76$ \\

Wall
&  & 
& $-2.35$ $(2.70)$ & $-0.87$
&  & 
& $0.40$ $(2.97)$ & $0.14$ \\

Building
&  & 
& $-2.89^{***}$ $(0.83)$ & $-3.47$
&  & 
& $-2.40^{**}$ $(0.91)$ & $-2.64$ \\

Sidewalk
&  & 
& $-8.03^{***}$ $(1.67)$ & $-4.81$
&  & 
& $-9.37^{***}$ $(1.81)$ & $-5.17$ \\

\hline
$\ln(\mathrm{PS})$
& $-0.67^{***}$ $(0.04)$ & $-16.00$
& $-0.69^{***}$ $(0.04)$ & $-16.30$
& $-0.67^{***}$ $(0.04)$ & $-15.07$
& $-0.68^{***}$ $(0.04)$ & $-15.14$ \\

\hline
\multicolumn{9}{l}{\textit{Model fit}} \\

\textit{n}
& \multicolumn{2}{c}{84,631}
& \multicolumn{2}{c}{84,631}
& \multicolumn{2}{c}{84,631}
& \multicolumn{2}{c}{84,631} \\

\textit{Events}
& \multicolumn{2}{c}{9,686}
& \multicolumn{2}{c}{9,686}
& \multicolumn{2}{c}{9,686}
& \multicolumn{2}{c}{9,686} \\
\textit{Concordance}
& \multicolumn{2}{c}{0.779 (0.003)}
& \multicolumn{2}{c}{0.782 (0.003)}
& \multicolumn{2}{c}{0.790 (0.003)}
& \multicolumn{2}{c}{0.795 (0.003)} \\
\textit{McFadden $R^2$}        & \multicolumn{2}{c}{0.190} & \multicolumn{2}{c}{0.193} & \multicolumn{2}{c}{0.219} & \multicolumn{2}{c}{0.222} \\

\textit{LRT\textsuperscript{1} (vs Null)}
& \multicolumn{2}{c}{6970 on 6 df}
& \multicolumn{2}{c}{7094 on 13 df}
& \multicolumn{2}{c}{8029 on 21 df}
& \multicolumn{2}{c}{8144 on 28 df} \\
\textit{LRT\textsuperscript{1} (vs Base)}         & \multicolumn{2}{c}{--}  & \multicolumn{2}{c}{$124^{***}$} & \multicolumn{2}{c}{$1059^{***}$} & \multicolumn{2}{c}{$1174^{***}$} \\
\textit{LRT\textsuperscript{1} (vs Infra)}        & \multicolumn{2}{c}{--}  & \multicolumn{2}{c}{--}  & \multicolumn{2}{c}{--}  & \multicolumn{2}{c}{$115^{***}$} \\
\hline
\end{tabular}%
}
\par\vspace{2mm}
\footnotesize
\textbf{Note:} \textsuperscript{1}Likelihood ratio test. $^{***}p<0.001$; $^{**}p<0.01$; $^{*}p<0.05$; $^{.}p<0.1$.
\end{table}

\textcolor{black}{
We estimated four conditional logit models using the R package \texttt{clogit} \citep{reid2014clogitL1} to evaluate the incremental contribution of different variable groups (Table~\ref{results-1234}). Model~1 includes only trip characteristics and elevation variables. Models~2 and~3 separately add SVI-derived street-level visual features and GIS-derived infrastructure variables. Model~4 includes all variables. The full model achieved a Concordance Index (C-index)\footnote{The C-index ranges from 0 to 1 and measures the model's ability to correctly rank the chosen route relative to non-chosen alternatives. Values above 0.5 indicate predictive performance better than random chance.} of 0.795 and McFadden's pseudo-$R^2$ of 0.22. Likelihood ratio tests showed that both SVI and infrastructure variables significantly improved model fit, although the contribution of infrastructure variables was substantially larger, consistent with previous studies \citep{xiao2025effects, lieu2026comparing, juarez2023cyclists}. Notably, the route length coefficient became negative only after infrastructure variables were introduced, suggesting that they capture important route characteristics omitted from simpler specifications. Adding SVI variables to Model~3 produced only minor changes in the estimated coefficients. To reduce the influence of sporadic or potentially noisy trajectories, only observed routes with a frequency of at least two were retained as chosen alternatives in the final hybrid choice set. This filtering had little influence on the estimated coefficients while producing the expected negative route length coefficient (Appendix Table~\ref{results-123}). In addition, a likelihood ratio test indicated that the PSL specification provided a significant improvement in model fit over the corresponding Multinomial Logit (MNL) model without the $\ln(\mathrm{PS})$ term (Appendix Table~\ref{results-mnl-psl}). Collectively, these results support the use of Model~4 as the preferred specification, and the following discussion therefore focuses on the full model.}

\textcolor{black}{
The negative path size coefficient, while seemingly counterintuitive, has also been reported in previous cyclist route choice studies \citep{scott2021route, lilasathapornkit2025cycling}. It indicates that routes with greater overlap among the alternatives were more likely to be chosen after controlling for other route attributes. When all observed routes were retained in the hybrid choice set, the path size coefficient became positive. After excluding low-frequency routes (frequency = 1), it became negative (Appendix Table~\ref{results-123}). This finding provides empirical support for the hypothesis that the estimated path size effect depends on the behavioral composition of the choice set \citep{lilasathapornkit2025cycling}.}

Route choice was negatively associated with route length, indicating that e-bike users tended to prefer shorter routes when other variables were held constant. This finding is consistent with most existing bike and e-bike route choice studies \citep{halldorsdottir2014efficiency,meister2024comparative,lukawska2023joint,meister2023route,ton2018evaluating,scott2021route}. The number of turns was also negatively associated with route choice, suggesting that e-bike users generally preferred routes with fewer directional changes.

\textcolor{black}{
For the road infrastructure variables, the proportions sum to one. Therefore, residential roads were omitted as the reference category. Major protected bike lanes showed the largest positive coefficient among all infrastructure types, exceeding both painted bike lanes and shared lanes on major roads. A similar pattern was observed for protected facilities on minor roads. In contrast, major roads without bike lanes were strongly negatively associated with route choice, whereas the corresponding effect on minor roads was weaker and only marginally significant. These findings are consistent with previous studies showing that protected bike lanes attract cyclists from alternative routes \citep{monsere2014lessons} and provide greater perceived comfort and safety than painted facilities \citep{mcneil2015influence}.}

\textcolor{black}{
The negative coefficients for footpaths, crossings, pedestrian zones, and shared paths suggest that e-bike riders generally avoid routes involving frequent interactions with pedestrians or motor vehicles. Crossings may be particularly undesirable because they interrupt riding continuity through more frequent traffic conflicts. This interpretation is further supported by the positive coefficient for off-street bike paths, which provide dedicated and uninterrupted riding environments.}

The model results indicated that slope still influenced e-bike route choice. However, only routes containing segments with slopes exceeding 10\% were significantly less likely to be chosen, whereas moderate slopes had weaker and non-significant effects. This suggests that electric assistance mitigates the deterrent effect of moderate uphill segments, while steep gradients remain undesirable because they increase physical effort, reduce riding comfort, and may compromise safety and travel efficiency. Similar findings have been reported in previous studies \citep{broach2012cyclists, yin2026pedal}. \textcolor{black}{Likewise, stairs were negatively associated with route choice, reflecting a preference for barrier-free routes.}

\textcolor{black}{
Finally, several street-level visual features were significantly associated with e-bike route choice. Among the vegetation variables, only tree coverage showed a positive effect. This finding is consistent with previous bicycle route choice studies \citep{park2019bicyclists, xiao2025effects, lieu2026comparing} and likely reflects the benefits of shade and thermal comfort. In contrast, other vegetation types were not significant, suggesting that trees provide more functional benefits for riders than low vegetation. The effects of sky visibility and building coverage are less straightforward and may partly reflect broader urban form and surrounding environmental characteristics. Sidewalk visibility was also negatively associated with route choice, reinforcing the infrastructure findings that riders prefer dedicated cycling facilities over pedestrian-oriented infrastructure.}

\subsection{Context dependence and interaction effects}

\begin{figure}[!t]
  \centering
  \includegraphics[width=1\textwidth]{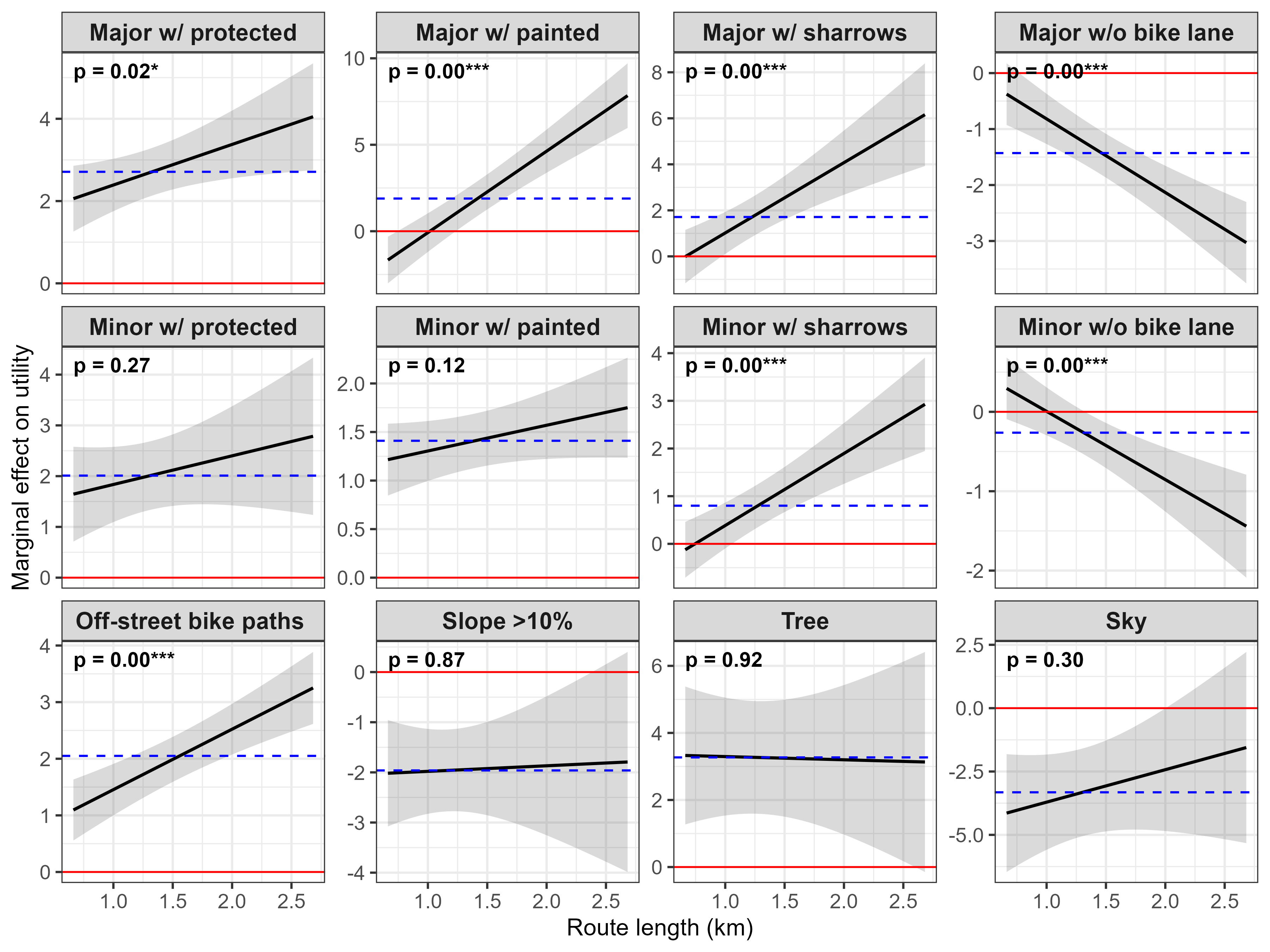}
  \caption{Interaction effects between route length and selected route attributes. Each panel shows the estimated marginal effect ($\beta_X+\beta_{XL}L$) of a route attribute on route utility. Shaded areas denote 95\% confidence intervals. The blue dashed line represents the estimate without the interaction term, and the red line indicates zero effect.}
  \label{var_inter}
\end{figure}

\textcolor{black}{
Although the main model estimates average preferences across all trips, the influence of route attributes may vary with trip context. As summarized by \citet{hallberg2021modelling}, e-bike users typically travel at higher speeds and cover longer distances than conventional cyclists. As trip length increases, the cumulative effects of route conditions become more pronounced, potentially altering the relative importance of different route attributes. We therefore examined whether the effects of selected infrastructure and street-level visual features vary with route length to assess the context dependence of route preferences.
}

\textcolor{black}{
For each selected attribute \(X\), we estimated a separate model including its interaction with route length \(L\). The marginal effect of \(X\) is therefore:
\[
\frac{\partial U}{\partial X}=\beta_X+\beta_{XL}L .
\]
Route length was evaluated over the observed 5th--95th percentile range to avoid extrapolation beyond the main body of the data.
}

\textcolor{black}{
Figure~\ref{var_inter} shows that bike facilities on major roads exhibited the strongest context dependence. The positive effects of protected bike lanes, painted bike lanes, and shared lanes all increased with route length, whereas major roads without cycling facilities became increasingly unattractive as route length increased. This suggests that riders place greater value on cycling facilities during longer trips, when comfort and continuity become more important. Although painted bike lanes and shared lanes were positively associated with route choice in the main model, their effects were relatively small for short and medium-length trips but became stronger as route length increased. In contrast, protected bike lanes maintained a consistently positive effect across the observed route length range. The interactions for slopes exceeding 10\% and tree coverage were not statistically significant, and their marginal effects remained nearly constant across route lengths.
}

\subsection{Effect sizes}

\begin{table}[!t]
\centering
\footnotesize
\renewcommand\arraystretch{1.2}
\caption{\textcolor{black}{Effect sizes for one-standard-deviation changes}}
\label{tab:effect_sd}
\vspace{4pt}

\begin{tabular}{lrrrrrc}
\hline
Variable & Coef. & SD & Coef.$\times$SD & OR$_{1\mathrm{SD}}$ & 95\% CI & Sig. \\
\hline

Turns                         & $-0.28$ & $6.68$  & $-1.88$ & $0.15$ & $(0.14,\ 0.16)$ & $***$ \\
Minor w/ painted              & $1.41$  & $25.72$ & $0.36$  & $1.44$ & $(1.34,\ 1.54)$ & $***$ \\
Crossings                     & $-8.92$ & $3.65$  & $-0.33$ & $0.72$ & $(0.69,\ 0.75)$ & $***$ \\
Length                        & $-0.48$ & $0.66$  & $-0.32$ & $0.73$ & $(0.59,\ 0.90)$ & $**$ \\
Major w/o bike lane           & $-1.43$ & $22.34$ & $-0.32$ & $0.73$ & $(0.67,\ 0.79)$ & $***$ \\
Off-street bike paths         & $2.05$  & $15.57$ & $0.32$  & $1.38$ & $(1.29,\ 1.46)$ & $***$ \\
Major w/ protected            & $2.71$  & $11.40$ & $0.31$  & $1.36$ & $(1.27,\ 1.46)$ & $***$ \\
Stairs                        & $-139.72$ & $0.16$ & $-0.22$ & $0.80$ & $(0.75,\ 0.85)$ & $***$ \\
Minor w/ protected            & $2.01$  & $10.61$ & $0.21$  & $1.24$ & $(1.15,\ 1.33)$ & $***$ \\
Major w/ painted              & $1.89$  & $10.36$ & $0.20$  & $1.22$ & $(1.09,\ 1.35)$ & $***$ \\
Slope $>$10\%                 & $-1.96$ & $9.10$  & $-0.18$ & $0.84$ & $(0.78,\ 0.90)$ & $***$ \\
Tree                          & $3.27$  & $5.44$  & $0.18$  & $1.19$ & $(1.09,\ 1.31)$ & $***$ \\
Sidewalk                      & $-9.37$ & $1.72$  & $-0.16$ & $0.85$ & $(0.80,\ 0.90)$ & $***$ \\
Building                      & $-2.40$ & $6.62$  & $-0.16$ & $0.85$ & $(0.76,\ 0.96)$ & $**$ \\
Major w/ sharrows             & $1.71$  & $7.87$  & $0.13$  & $1.14$ & $(1.07,\ 1.22)$ & $***$ \\
Shared paths                  & $-2.26$ & $5.84$  & $-0.13$ & $0.88$ & $(0.84,\ 0.92)$ & $***$ \\
Sky                           & $-3.32$ & $3.97$  & $-0.13$ & $0.88$ & $(0.82,\ 0.94)$ & $***$ \\
Footpath                      & $-1.62$ & $7.18$  & $-0.12$ & $0.89$ & $(0.85,\ 0.93)$ & $***$ \\
Minor w/ sharrows             & $0.80$  & $12.73$ & $0.10$  & $1.11$ & $(1.05,\ 1.17)$ & $***$ \\
Grass                         & $5.82$  & $1.21$  & $0.07$  & $1.07$ & $(0.98,\ 1.17)$ & \\
Minor w/o bike lane           & $-0.26$ & $22.32$ & $-0.06$ & $0.94$ & $(0.89,\ 1.00)$ & $. $ \\
Pedestrian zones              & $-6.45$ & $0.81$  & $-0.05$ & $0.95$ & $(0.91,\ 0.99)$ & $*$ \\
Plant                         & $5.76$  & $0.64$  & $0.04$  & $1.04$ & $(0.97,\ 1.11)$ & \\
Slope 2\%-6\%                 & $0.26$  & $10.02$ & $0.03$  & $1.03$ & $(0.98,\ 1.08)$ & \\
Slope 6\%-10\%                & $-0.32$ & $6.52$  & $-0.02$ & $0.98$ & $(0.94,\ 1.02)$ & \\
Living streets                & $0.49$  & $0.87$  & $0.00$  & $1.00$ & $(0.98,\ 1.03)$ & \\
Wall                          & $0.40$  & $0.91$  & $0.00$  & $1.00$ & $(0.95,\ 1.06)$ & \\

\hline
\end{tabular}%

\par\vspace{2mm}
\footnotesize
\raggedright
\textbf{Note:} For variables measured as proportions, SD is reported in percentage points. $^{***}p<0.001$; $^{**}p<0.01$; $^{*}p<0.05$; $^{.}p<0.1$.
\end{table}

\textcolor{black}{
Because the explanatory variables differ not only in measurement scale but also in their observed variability, the estimated coefficients are not directly comparable in terms of their practical importance. To facilitate comparison, Table~\ref{tab:effect_sd} reports the effect associated with a one-standard-deviation increase in each variable. Specifically, Coef.$\times$SD represents the corresponding change in utility and is used to compare effect sizes across variables, while OR$_{1\mathrm{SD}}$ gives the corresponding change in route choice odds. For variables measured as proportions, the standard deviation is reported in percentage points for interpretability, whereas Coef.$\times$SD, OR$_{1\mathrm{SD}}$, and the corresponding confidence intervals are calculated using the original 0--1 scale.}

\textcolor{black}{
Among the infrastructure variables, minor roads with painted bike lanes, off-street bike paths, and major roads with protected bike lanes exhibited the largest positive effect sizes, whereas major roads without bike lanes and crossings exhibited the largest negative effect sizes. The magnitudes of these effects were comparable to that of route length. Among the streetscape features, trees exhibited the largest positive effect size.}

\section{Discussion}

This section discusses the implications of the findings for understanding e-bike route choice behavior and designing e-bike-friendly cycling infrastructure. It also compares the observed route preferences with findings from previous studies on bicycles and e-scooters.

\subsection{Implications for cycling infrastructure planning and design}
\textcolor{black}{
On average, preferred e-bike routes reflected two overarching themes: dedicated riding space that minimizes conflicts with both motor vehicles and pedestrians, and route continuity.}

\textcolor{black}{
Low-conflict riding environments appeared to be a defining characteristic of preferred e-bike routes. Protected bike lanes were substantially more attractive than roads without cycling facilities, and this gap was both wider on major roads than on minor roads and increased with route length, suggesting that the benefit of separation from motor traffic increased with both the intensity and duration of vehicle exposure. Painted bike lanes, which provide designated space without physical separation, remained considerably more attractive than undesignated roads. At the same time, the negative associations observed for footpaths, pedestrian zones, shared paths, and sidewalks indicate that e-bike users also tended to avoid environments with greater potential conflicts with pedestrians. Together, these patterns point to a consistent preference for infrastructure that clearly distinguishes bicycle traffic from both motor vehicles and pedestrians. Route continuity emerged as a second defining characteristic. The negative associations for crossings, turns, and stairs indicate that riders preferred routes with fewer interruptions, prioritizing a smooth and unbroken riding experience.}

\textcolor{black}{
The standardized effect size analysis further reinforces these two themes. Within the observed variability of the existing road network, attributes representing dedicated cycling space and route continuity consistently exhibited the largest effect sizes, indicating that these characteristics offer the greatest potential for influencing route preferences among the infrastructure attributes examined in this study.}

\subsection{Comparisons with bicycle and e-scooter route choice behavior}
\textcolor{black}{
To place the findings in context, we relate the observed e-bike route preferences to prior evidence for bicycles and e-scooters. Although these modes share similar street networks, differences in vehicle characteristics and typical travel contexts suggest that route choice sensitivities may vary across modes.}

\textcolor{black}{
Many route preference patterns commonly reported for bicycle riders were also observed in our e-bike sample, including preferences for higher-quality cycling facilities and positive associations with tree coverage, suggesting broadly similar route preferences across the two modes. However, several notable differences also emerged. Existing bicycle studies have generally reported stronger sensitivity to steep slopes and greater aversion to arterial roads with painted bike lanes \citep{lilasathapornkit2025cycling,chavis2021bikes,meister2023route}. In our e-bike sample, only slopes exceeding 10\% were significantly associated with route avoidance, suggesting that electric assistance mitigates, but does not eliminate, the deterrent effect of steep terrain. A second difference concerns painted bike lanes on major roads. Unlike the stronger aversion reported in the bicycle literature, painted bike lanes remained positively associated with route choice in our e-bike sample, consistent with evidence that e-bike riders tend to report greater subjective safety than conventional cyclists \citep{arning2023review}. Motor assistance may also enable riders to maintain more consistent speeds under mixed traffic conditions, making painted facilities more acceptable. Major roads also appeared to be more attractive when bicycle facilities were present, consistent with evidence that e-bikes are more commonly used on major roads where their speed advantage over conventional bicycles is more readily realized \citep{khavarian2024bike,langford2015risky}.}

\textcolor{black}{
The evidence for e-scooters discussed here is drawn from a Washington, DC study using a comparable modeling framework \citep{qian_2026_20836439}. Overall, route choice patterns for road types, bicycle facilities and greenery were also broadly similar across the two electrically assisted modes. However, compared with the e-bike results, e-scooter riders appeared more willing to use sidewalks in the absence of bicycle facilities and exhibited a stronger preference for minor and residential roads. These differences may reflect both vehicle characteristics and travel context. Compared with e-bikes, e-scooters generally provide lower riding stability, making interactions with lane changes, overtaking vehicles, and heavy traffic more challenging. In contrast, e-bikes have higher cruising speeds and are more commonly used for commuting, which may make major roads with dedicated bicycle facilities relatively more attractive. Sensitivity to elevation also differed between the two modes. In the e-scooter study, slope-related variables were largely statistically insignificant, indicating limited sensitivity to elevation changes. In contrast, only steep slopes ($>$10\%) were significantly avoided in our e-bike model. This difference may partly reflect the fact that shared e-bikes in Washington, DC are primarily pedal-assist, requiring riders to continue providing human effort during climbing, whereas e-scooters rely on fully motorized propulsion with no pedaling involvement.}

\section{Conclusion}

\textcolor{black}{
This study examined shared e-bike route choice in the urban environment of Washington, DC, using GPS-based revealed preference data and a PSL model with a hybrid choice set consisting of observed routes and their corresponding shortest paths. The results indicate that e-bike riders generally preferred routes that combined dedicated cycling space with continuous travel conditions, thereby minimizing conflicts with both motor vehicles and pedestrians. Roadway hierarchy and bicycle facility type jointly shaped route preferences, with the presence of bicycle facilities having a much larger influence on route choice along major roads than along minor roads. Furthermore, longer trips exhibit stronger preferences for cycling infrastructure, indicating that the effectiveness of cycling infrastructure is jointly determined by the intensity and cumulative duration of riders' exposure to surrounding traffic.}

\textcolor{black}{
Beyond conventional GIS-based infrastructure variables, this study incorporated street-level visual features extracted from SVI to capture street-level visual context. These variables significantly improved model performance and provided additional insights into e-bike route preferences, although their contributions to model fit and effect sizes were generally smaller than those of infrastructure variables. Methodologically, the results demonstrate the value of integrating computer vision-based streetscape measures into behavioral route choice models, providing a practical approach for incorporating high-resolution street context into micromobility research.}

\textcolor{black}{
From a cross-mode perspective, the findings suggest that e-bike route choice shares many characteristics with both bicycles and e-scooters, particularly preferences for high-quality cycling infrastructure, although e-bike riders may be somewhat more tolerant of bicycle facilities without physical separation.}

Several limitations of this study should be acknowledged. \textcolor{black}{First, bicycle infrastructure is not randomly distributed across the road network, introducing potential endogeneity that we do not fully address in this study. Hence, we have interpreted the results mostly as associative rather than causal.} Second, the model does not account for individual-level demographic or socioeconomic characteristics due to data constraints, which means unobserved preference heterogeneity is not properly addressed. Future research could address this limitation by integrating the collection of GPS-based trip trajectory data with user surveys to create more comprehensive datasets that support model structures capable of more accurately capturing behavioral heterogeneity. \textcolor{black}{Third, street-level visual features were extracted from single-time SVI using a pre-trained computer vision model without task-specific fine-tuning. This may introduce measurement error and cannot capture temporal variations in dynamic street elements. Future research could address these limitations through local model fine-tuning and multi-temporal street-level imagery.} Finally, this analysis focuses on Washington, DC and its surrounding areas. Additional studies across different urban forms and transportation contexts are needed to assess the generalizability of these findings.

\section{Acknowledgments}
We are grateful for the funding support from U.S. National Science Foundation (Award \#2425029). ChatGPT was used to improve the readability and language of the work.

\bibliographystyle{elsarticle-harv}
\biboptions{semicolon,round,sort,authoryear}
\bibliography{sample}

\newpage
\appendix

\section{\textcolor{black}{Sensitivity analysis of observed route inclusion criteria}}
\setcounter{table}{0}
\begin{table}[H]
\centering
\footnotesize
\renewcommand\arraystretch{1.2}
\caption{Path size logit model results by minimum route frequency}
\label{results-123}
\vspace{4pt}

\resizebox{\textwidth}{!}{%
\begin{tabular}{lcccccc}
\hline
& \multicolumn{2}{c}{Model (Freq $\geq$ 1)}
& \multicolumn{2}{c}{Model (Freq $\geq$ 2)}
& \multicolumn{2}{c}{Model (Freq $\geq$ 3)} \\
\cmidrule(lr){2-3}\cmidrule(lr){4-5}\cmidrule(lr){6-7}
Variable & Coef. (SE) & z & Coef. (SE) & z & Coef. (SE) & z \\
\hline

\multicolumn{7}{l}{\textbf{Trip characteristics}} \\

Length
& $0.00$ $(0.11)$ & $0.04$
& $-0.48^{**}$ $(0.16)$ & $-3.00$
& $-0.58^{**}$ $(0.18)$ & $-3.20$ \\
Turns
& $-0.19^{***}$ $(0.00)$ & $-58.02$
& $-0.28^{***}$ $(0.00)$ & $-59.52$
& $-0.31^{***}$ $(0.01)$ & $-56.49$ \\
\hline
\multicolumn{7}{l}{\textbf{Elevation}} \\
Slope 2\%-6\%
& $0.09$ $(0.18)$ & $0.51$
& $0.26$ $(0.24)$ & $1.08$
& $0.03$ $(0.27)$ & $0.12$ \\

Slope 6\%-10\%
& $-0.61^{*}$ $(0.27)$ & $-2.26$
& $-0.32$ $(0.35)$ & $-0.92$
& $-0.31$ $(0.39)$ & $-0.80$ \\

Slope $>$10\%
& $-1.44^{***}$ $(0.31)$ & $-4.67$
& $-1.96^{***}$ $(0.42)$ & $-4.71$
& $-1.98^{***}$ $(0.47)$ & $-4.20$ \\

\hline
\multicolumn{7}{l}{\textbf{Road infrastructure}} \\
Residential (ref.)
& -- & --
& -- & --
& -- & -- \\

Major w/ protected
& $1.46^{***}$ $(0.21)$ & $7.00$
& $2.71^{***}$ $(0.30)$ & $9.13$
& $3.39^{***}$ $(0.34)$ & $9.92$ \\
Major w/ painted
& $1.22^{***}$ $(0.36)$ & $3.34$
& $1.89^{***}$ $(0.52)$ & $3.61$
& $2.52^{***}$ $(0.59)$ & $4.27$ \\
Major w/ sharrows
& $1.03^{***}$ $(0.31)$ & $3.36$
& $1.71^{***}$ $(0.42)$ & $4.05$
& $2.10^{***}$ $(0.49)$ & $4.28$ \\
Major w/o bike lane
& $-0.73^{***}$ $(0.14)$ & $-5.20$
& $-1.43^{***}$ $(0.20)$ & $-7.25$
& $-1.58^{***}$ $(0.22)$ & $-7.08$ \\
Minor w/ protected
& $1.25^{***}$ $(0.25)$ & $5.08$
& $2.01^{***}$ $(0.34)$ & $5.84$
& $2.37^{***}$ $(0.39)$ & $6.05$ \\
Minor w/ painted
& $0.64^{***}$ $(0.10)$ & $6.35$
& $1.41^{***}$ $(0.14)$ & $9.84$
& $1.63^{***}$ $(0.16)$ & $10.00$ \\
Minor w/ sharrows
& $0.68^{***}$ $(0.17)$ & $3.96$
& $0.80^{***}$ $(0.23)$ & $3.48$
& $1.12^{***}$ $(0.27)$ & $4.16$ \\
Minor w/o bike lane
& $-0.24^{*}$ $(0.10)$ & $-2.34$
& $-0.26^{.}$ $(0.14)$ & $-1.90$
& $-0.20$ $(0.16)$ & $-1.31$ \\
Crossings
& $-7.35^{***}$ $(0.45)$ & $-16.25$
& $-8.92^{***}$ $(0.59)$ & $-15.24$
& $-9.67^{***}$ $(0.65)$ & $-14.93$ \\
Footpath
& $-1.52^{***}$ $(0.27)$ & $-5.65$
& $-1.62^{***}$ $(0.34)$ & $-4.76$
& $-2.01^{***}$ $(0.38)$ & $-5.25$ \\
Off-street bike paths
& $1.14^{***}$ $(0.15)$ & $7.79$
& $2.05^{***}$ $(0.20)$ & $10.24$
& $2.37^{***}$ $(0.23)$ & $10.46$ \\
Pedestrian zones
& $-2.51$ $(1.93)$ & $-1.30$
& $-6.45^{*}$ $(2.71)$ & $-2.38$
& $-6.63^{*}$ $(3.02)$ & $-2.19$ \\
Living streets
& $1.17$ $(1.12)$ & $1.04$
& $0.49$ $(1.49)$ & $0.33$
& $0.87$ $(1.62)$ & $0.54$ \\
Shared paths
& $-0.55^{*}$ $(0.27)$ & $-2.04$
& $-2.26^{***}$ $(0.40)$ & $-5.59$
& $-3.00^{***}$ $(0.46)$ & $-6.52$ \\
Stairs
& $-95.23^{***}$ $(14.63)$ & $-6.51$
& $-139.72^{***}$ $(21.05)$ & $-6.64$
& $-169.00^{***}$ $(25.30)$ & $-6.68$ \\
\hline
\multicolumn{7}{l}{\textbf{Street-level visual features}} \\

Tree
& $2.85^{***}$ $(0.63)$ & $4.56$
& $3.27^{***}$ $(0.86)$ & $3.82$
& $4.41^{***}$ $(0.98)$ & $4.48$ \\

Plant
& $1.51$ $(3.71)$ & $0.41$
& $5.76$ $(5.11)$ & $1.13$
& $2.62$ $(5.78)$ & $0.45$ \\

Grass
& $1.82$ $(2.76)$ & $0.66$
& $5.82$ $(3.79)$ & $1.54$
& $5.38$ $(4.32)$ & $1.25$ \\

Sky
& $-2.07^{**}$ $(0.65)$ & $-3.18$
& $-3.32^{***}$ $(0.88)$ & $-3.76$
& $-3.31^{***}$ $(1.00)$ & $-3.30$ \\

Wall
& $1.43$ $(2.20)$ & $0.65$
& $0.40$ $(2.97)$ & $0.14$
& $3.16$ $(3.32)$ & $0.95$ \\

Building
& $-1.44^{*}$ $(0.66)$ & $-2.17$
& $-2.40^{**}$ $(0.91)$ & $-2.64$
& $-2.48^{*}$ $(1.03)$ & $-2.41$ \\

Sidewalk
& $-1.92$ $(1.32)$ & $-1.46$
& $-9.37^{***}$ $(1.81)$ & $-5.17$
& $-9.80^{***}$ $(2.04)$ & $-4.80$ \\

\hline
$\ln(\mathrm{PS})$
& $0.09^{**}$ $(0.03)$ & $2.76$
& $-0.68^{***}$ $(0.04)$ & $-15.14$
& $-0.88^{***}$ $(0.05)$ & $-17.09$ \\

\hline
\multicolumn{7}{l}{\textit{Model fit}} \\

\textit{n}
& \multicolumn{2}{c}{129,586}
& \multicolumn{2}{c}{84,631}
& \multicolumn{2}{c}{69,521} \\

\textit{Events}
& \multicolumn{2}{c}{13,118}
& \multicolumn{2}{c}{9,686}
& \multicolumn{2}{c}{8,524} \\

\textit{Concordance}
& \multicolumn{2}{c}{0.743 (0.003)}
& \multicolumn{2}{c}{0.795 (0.003)}
& \multicolumn{2}{c}{0.808 (0.003)} \\

\textit{McFadden $R^2$}        & \multicolumn{2}{c}{0.137} & \multicolumn{2}{c}{0.222}& \multicolumn{2}{c}{0.253}\\
\hline
\end{tabular}%
}
\par\vspace{2mm}
\footnotesize
\textbf{Note:} $^{***}p<0.001$; $^{**}p<0.01$; $^{*}p<0.05$; $^{.}p<0.1$.
\end{table}

This appendix presents a sensitivity analysis of the route expansion strategy used to construct the route choice dataset. Alternatives with a frequency of one represent routes that were observed only once, even after OD aggregation, and may therefore reflect infrequent or idiosyncratic route choices. To evaluate the robustness of the model, we consider three expansion strategies, denoted as Freq1--Freq3. Freq1 retains all observed alternatives, whereas Freq3 progressively excludes low-frequency alternatives and focuses on dominant routes within each OD pair. The resulting model estimates are compared to assess the sensitivity of the findings to the treatment of infrequently observed alternatives.

\section{\textcolor{black}{PSL vs MNL}}
\setcounter{table}{0}
\begin{table}[H]
\centering
\scriptsize
\renewcommand\arraystretch{1.2}
\caption{Comparison of multinomial logit and path size logit model results}
\label{results-mnl-psl}
\vspace{4pt}
\begin{tabular}{lcccc}
\hline
& \multicolumn{2}{c}{MNL}
& \multicolumn{2}{c}{PSL} \\
\cmidrule(lr){2-3}\cmidrule(lr){4-5}
Variable & Coef. (SE) & z & Coef. (SE) & z \\
\hline

\multicolumn{5}{l}{\textbf{Trip characteristics}} \\

Length
& $-1.45^{***}$ $(0.15)$ & $-9.83$
& $-0.48^{**}$ $(0.16)$ & $-3.00$ \\

Turns
& $-0.27^{***}$ $(0.00)$ & $-58.23$
& $-0.28^{***}$ $(0.00)$ & $-59.52$ \\

\hline
\multicolumn{5}{l}{\textbf{Elevation}} \\

Slope 2\%-6\%
& $0.30$ $(0.24)$ & $1.29$
& $0.26$ $(0.24)$ & $1.08$ \\

Slope 6\%-10\%
& $-0.17$ $(0.34)$ & $-0.51$
& $-0.32$ $(0.35)$ & $-0.92$ \\

Slope $>$10\%
& $-1.78^{***}$ $(0.40)$ & $-4.43$
& $-1.96^{***}$ $(0.42)$ & $-4.71$ \\

\hline
\multicolumn{5}{l}{\textbf{Road infrastructure}} \\

Residential (ref.)
& -- & --
& -- & -- \\

Major w/ protected
& $2.85^{***}$ $(0.28)$ & $10.02$
& $2.71^{***}$ $(0.30)$ & $9.13$ \\

Major w/ painted
& $2.19^{***}$ $(0.50)$ & $4.35$
& $1.89^{***}$ $(0.52)$ & $3.61$ \\

Major w/ sharrows
& $1.76^{***}$ $(0.40)$ & $4.34$
& $1.71^{***}$ $(0.42)$ & $4.05$ \\

Major w/o bike lane
& $-1.30^{***}$ $(0.19)$ & $-6.93$
& $-1.43^{***}$ $(0.20)$ & $-7.25$ \\

Minor w/ protected
& $2.01^{***}$ $(0.33)$ & $6.10$
& $2.01^{***}$ $(0.34)$ & $5.84$ \\

Minor w/ painted
& $1.45^{***}$ $(0.14)$ & $10.56$
& $1.41^{***}$ $(0.14)$ & $9.84$ \\

Minor w/ sharrows
& $0.96^{***}$ $(0.22)$ & $4.30$
& $0.80^{***}$ $(0.23)$ & $3.48$ \\

Minor w/o bike lane
& $-0.17$ $(0.13)$ & $-1.31$
& $-0.26^{.}$ $(0.14)$ & $-1.90$ \\

Crossings
& $-8.77^{***}$ $(0.57)$ & $-15.36$
& $-8.92^{***}$ $(0.59)$ & $-15.24$ \\

Footpath
& $-1.41^{***}$ $(0.33)$ & $-4.27$
& $-1.62^{***}$ $(0.34)$ & $-4.76$ \\

Off-street bike paths
& $2.02^{***}$ $(0.19)$ & $10.43$
& $2.05^{***}$ $(0.20)$ & $10.24$ \\

Pedestrian zones
& $-5.41^{*}$ $(2.65)$ & $-2.04$
& $-6.45^{*}$ $(2.71)$ & $-2.38$ \\

Living streets
& $0.58$ $(1.47)$ & $0.40$
& $0.49$ $(1.49)$ & $0.33$ \\

Shared paths
& $-2.40^{***}$ $(0.40)$ & $-6.06$
& $-2.26^{***}$ $(0.40)$ & $-5.59$ \\

Stairs
& $-136.60^{***}$ $(20.93)$ & $-6.53$
& $-139.72^{***}$ $(21.05)$ & $-6.64$ \\

\hline
\multicolumn{5}{l}{\textbf{Street-level visual features}} \\

Tree
& $3.17^{***}$ $(0.82)$ & $3.86$
& $3.27^{***}$ $(0.86)$ & $3.82$ \\

Plant
& $2.66$ $(4.93)$ & $0.54$
& $5.76$ $(5.11)$ & $1.13$ \\

Grass
& $6.59^{.}$ $(3.66)$ & $1.80$
& $5.82$ $(3.79)$ & $1.54$ \\

Sky
& $-3.27^{***}$ $(0.85)$ & $-3.85$
& $-3.32^{***}$ $(0.88)$ & $-3.76$ \\

Wall
& $1.48$ $(2.91)$ & $0.51$
& $0.40$ $(2.97)$ & $0.14$ \\

Building
& $-2.30^{**}$ $(0.88)$ & $-2.63$
& $-2.40^{**}$ $(0.91)$ & $-2.64$ \\

Sidewalk
& $-8.60^{***}$ $(1.75)$ & $-4.92$
& $-9.37^{***}$ $(1.81)$ & $-5.17$ \\

\hline
$\ln(\mathrm{PS})$
& -- & --
& $-0.68^{***}$ $(0.04)$ & $-15.14$ \\

\hline
\multicolumn{5}{l}{\textit{Model fit}} \\

\textit{n}
& \multicolumn{2}{c}{84,631}
& \multicolumn{2}{c}{84,631} \\

\textit{Events}
& \multicolumn{2}{c}{9,686}
& \multicolumn{2}{c}{9,686} \\

\textit{Concordance}
& \multicolumn{2}{c}{0.789 (0.003)}
& \multicolumn{2}{c}{0.795 (0.003)} \\
\textit{McFadden $R^2$}        & \multicolumn{2}{c}{0.215} & \multicolumn{2}{c}{0.222}\\
\textit{LRT\textsuperscript{1} (vs Null)}
& \multicolumn{2}{c}{7906 on 27 df}
& \multicolumn{2}{c}{8144 on 28 df} \\
\textit{LRT\textsuperscript{1} (vs MNL)}
& \multicolumn{2}{c}{--}
& \multicolumn{2}{c}{$238^{***}$} \\
\hline
\end{tabular}%
\par\vspace{2mm}
\footnotesize
\textbf{Note:} \textsuperscript{1}Likelihood ratio test. $^{***}p<0.001$; $^{**}p<0.01$; $^{*}p<0.05$; $^{.}p<0.1$.
\end{table}

\end{document}